\documentclass[unnumsec,webpdf,contemporary,large]{main}%

\theoremstyle{thmstyleone}%
\theoremstyle{thmstyletwo}%
\theoremstyle{thmstylethree}%

\usepackage{xspace}
\usepackage{amsmath} 
\usepackage{hyperref} 
\usepackage{graphicx} 
\usepackage{amssymb}
\usepackage{forest}
\usepackage{tikz}
\usepackage{tikz-qtree}
\usepackage{dsfont} 
\usepackage{fancyvrb}
\usepackage{xcolor} 
\definecolor{mycolor}{RGB}{255,69,0}

\newcommand{\nts}{\ensuremath{\text{\it nt}}\xspace}

\newcommand{\notes}[1]{}

\newcommand{\ith}[1]{\ensuremath{i^{{th}}}}

\newcount\permx
\newcount\permy
\def\permdot#1#2{
\permx=#1 \advance\permx by-1
\permy=#2 \advance\permy by-1
\psframe[fillcolor=black, fillstyle=solid]
(\permx,\permy)(#1, #2)
}

\newcommand{\argmin}{\operatornamewithlimits{\mathbf{argmin}}}

\newcommand{\vecx}{\ensuremath{\mathbf{x}}\xspace}

\newcommand{\vecy}{\ensuremath{\mathbf{y}}\xspace}

\newcommand{\smallnt}[1]{\ensuremath{_{\mbox{\tiny PP}}}\xspace}

\iffalse

\else

\fi

\newcommand{\smallurl}[1]{{\scriptsize \url{#1}}}

\newcommand{\leftb}{\ensuremath{\text{\tt (}}\xspace}
\newcommand{\rightb}{\ensuremath{\text{\tt )}}\xspace}

\newcommand{\mydot}{\ensuremath{\text{\tt .}}\xspace}

\newcommand{\linearfold}{{LinearFold}\xspace}

\newcommand{\nucA}{\ensuremath{\text{\sc a}}}
\newcommand{\nucU}{\ensuremath{\text{\sc u}}}
\newcommand{\nucC}{\ensuremath{\text{\sc c}}}
\newcommand{\nucG}{\ensuremath{\text{\sc g}}}
\newcommand{\gap}{\ensuremath{\text{\sc --}}}

\newcommand{\panel}[1]{\large \sf {#1}}

\newcommand{\lturbofold}{{LinearTurboFold}\xspace}

\newcommand{\turbofold}{{TurboFold}\xspace}
\newcommand{\turbofoldii}{{TurboFold II}\xspace}

\newcommand{\linearalignment}{{LinearAlignment}\xspace}

\newcommand{\locarna}{{LocARNA}\xspace}

\newcommand{\scarna}{{SCARNA}\xspace}

\newcommand{\rnaalifold}{{RNAalifold}\xspace}

\newcommand{\parts}{{PARTS}\xspace}

\newcommand{\lso}{{LinearSankoff}\xspace}
\newcommand{\sankoff}{{Sankoff}\xspace}
\newcommand{\dynalign}{{Dynalign}\xspace}
\newcommand{\dynalignii}{{Dynalign II}\xspace}
\newcommand{\foldalign}{{FoldAlign}\xspace}

\newcommand{\mafft}{{\text{MAFFT}}\xspace}
\newcommand{\rnastralign}{{\text{RNAStralign}}\xspace}
\newcommand{\multilign}{{\text{Multilign}}\xspace}

\begin{document}

\journaltitle{}
\DOI{DOI HERE}
\copyrightyear{}
\pubyear{}
\access{Advance Access Publication Date: Day Month Year}
\appnotes{Paper}

\firstpage{1}


\title[\lso]{\lso: Linear-time Simultaneous Folding and Alignment of RNA Homologs}

\author[1]{Sizhen Li}
\author[1]{Ning Dai}
\author[1,2]{He Zhang}
\author[1]{Apoorv Malik}
\author[3,4,5]{David H.~Mathews}
\author[1,$\ast$]{Liang Huang} 

\authormark{Author Name et al.}

\address[1]{\orgdiv{School of Electrical Engineering \& Computer Science}, \orgname{ Oregon State University}, \orgaddress{\street{Corvallis}, \postcode{97330}, \state{OR}, \country{USA}}}
\address[2]{\orgname{Baidu Research USA}, \orgaddress{\street{Sunnyvale}, \postcode{94089}, \state{CA}, \country{USA}}}
\address[3]{\orgdiv{Dept. of Biochemistry \& Biophysics}}
\address[4]{\orgdiv{Center for RNA Biology}}
\address[5]{\orgdiv{Dept. of Biostatistics \& Computational Biology}, \orgname{ University of Rochester Medical Center}, \orgaddress{\street{Rochester}, \postcode{14642}, \state{NY}, \country{14642}}}

\corresp[$\ast$]{Corresponding author. \href{liang.huang.sh@gmail.com}{liang.huang.sh@gmail.com}}




\abstract{The classical Sankoff algorithm for the simultaneous folding and alignment of homologous
RNA  sequences is highly influential, but it suffers from two major limitations in efficiency and modeling power.
First, it takes $O(n^6)$ for two sequences where $n$ is the average sequence length.
Most implementations and variations reduce the runtime to $O(n^3)$ by restricting the  alignment search space,
but this is still too slow for long sequences such as full-length viral genomes.
On the other hand, the Sankoff algorithm 
and all its existing implementations use a rather simplistic alignment model, 
which can result in poor alignment accuracy.
To address these problems, 
we propose \lso, which seamlessly integrates the original \sankoff algorithm with a powerful Hidden Markov Model-based alignment module. 
This extension substantially improves alignment quality,
which in turn benefits secondary structure prediction quality, confirmed over a diverse set of RNA families.
\lso also applies beam search heuristics and the A$^{\star}$ algorithm to achieve that runtime scales linearly with sequence length. 
\lso is the first linear-time algorithm for simultaneous folding and alignment,
and the first such algorithm to scale to coronavirus genomes ($n \simeq 30,000 \nts$).
It only takes 10 minutes for a  pair of SARS-CoV-2 and SARS-related genomes,
and outperforms previous work at identifying crucial conserved structures between the two genomes.}

\keywords{simultaneously folding and alignment, conserved secondary structure, structural alignment\\
{\bf Availability:} Code: \url{https://github.com/LinearFold/LinearSankoff}; Web Server: \url{http://www.linearfold.org/linearsankoff}
}

\maketitle

\section{Introduction}\label{sec1}

Many RNAs are involved in multiple cellular processes ~\cite{Eddy:2001, Doudna+Cech:2002},
whose functions highly rely on their conserved structures. 
Therefore, there is a need to develop fast and accurate algorithms
for conserved structure prediction over RNA homologs.

To automate comparative analysis, 
\sankoff~\cite{Sankoff:1985} pioneered an algorithm to simultaneously fold and align homologous sequences. 
However, it takes $O(n^{6})$ time for just two sequences with the average sequence length $n$,
and $O(n^{3k})$ time for $k$ sequences in general.
To make it feasible,
several 
implementations of the \sankoff algorithm~\cite{mathews2002dynalign,harmanci:+2007,fu2014dynalign,havgaard2007fast,do+:2005,will2007inferring,tabei2006scarna,harmanci2008parts}.
reduce the runtime to $O(n^3m^3)$ via banding the alignment region with a fixed width ($m$), 
which shrinks the alignment search space from $O(n^2)$ to $O(nm)$.
Some of these tools, including \parts~\cite{harmanci2008parts}, \locarna~\cite{will2007inferring} and \scarna~\cite{Tabei+:2008},
also simplified the energy model using base-pairing probabilities.
However, 
this cubic runtime is still intractable for long sequences such as full-length viral genomes.

Besides the intractability, there is yet another important limitation in the Sankoff framework and its implementations,
where the alignment module is overly simplistic which scores matches, mismatches, and gaps independently of each other (e.g., using the classical Needleman–Wunsch~\cite{needleman1970general} alignment).
For example, the original \sankoff algorithm and \dynalign~\cite{harmanci:+2007} only
include gap penalty, while \locarna and \foldalign also include mismatch matrices.
By contrast, the Hidden Markov Model (HMM) has been well studied and applied to align RNA sequences~\cite{durbin1998biological,Harmanci+:2011} which
score each alignment step (match, mismatch, or gap) depending on the previous step, therefore extending a gap is treated differently from starting a gap.

To address these problems,
we propose \lso,
which extends the original \sankoff algorithm (with full energy model) by incorporating a more powerful, HMM-based alignment model. 
This integration between Sankoff and HMM requires non-trivial generalizations to the original Sankoff-style dynamic programming algorithm.
As a result, all existing variants of Sankoff are simplified versions of \lso in terms of either or both folding and alignment models. 
To make it efficient, 
we generalize the beam pruning technique 
of \linearfold~\cite{huang+:2019} 
from single-sequence folding to homologous folding 
to make the \lso runtime scale linearly with the sum of sequence lengths.
More interestingly, \lso also applies the A$^{*}$ algorithm with admissible heuristics ~\cite{hart1968formal} together with beam pruning to further speed up the search. 

We make the following contributions:
\begin{itemize}
\item We provide the first rigorous formulation of simultaneous folding and alignment
using synchronous context free grammars borrowed from computational linguistics.
\item We integrate \sankoff with an HMM-based alignment model, which not only improves alignment quality but also in turn benefits folding quality, and generalize the \sankoff-style dynamic programming to keep track of HMM states.
\item We extend the beam search heuristic from single-sequence folding to joint folding to achieve linear runtime, and further apply the A* algorithm to speed up the search.
\item Overall, \lso achieves higher secondary structure prediction and alignment accuracies than three baseline models (\linearfold, \dynalign and \lturbofold).  
\end{itemize}

\section{Formulation and Modeling}\label{sec2}

\begin{table}
\centering
\begin{tabular}{c|c}
language & RNA\\
\hline
single-sentence parsing & single-sequence folding\\
CKY $O(n^3)$ & Nussinov/Zuker $O(n^3)$\\
context-free grammar (CFG) & CFG\\
\hline
synchronous parsing & homologous folding\\
(joint parsing \& alignment) & (joint folding \& alignment)\\
Wu \cite{wu:1997} $O(n^6)$ & Sankoff \cite{Sankoff:1985} $O(n^6)$\\
synchronous CFG & synchronous CFG
\end{tabular}
\smallskip
\caption{Correspondence between natural language parsing and RNA folding.
While the correspondence between single-sentence parsing and singe-sequence folding
is well-known, our work is the first to establish the connection between synchronous parsing and homologous folding. This leads to our borrowing of synchronous context-free grammar from the former to the latter.\label{tab:scfg}}
\end{table}

\subsection{Synchronous Context Free Grammar Formulation}\label{subsec1}
While there have been many variants of the Sankoff algorithm in the literature
\cite{Sankoff:1985,mathews2002dynalign,fu2014dynalign,havgaard2007fast,do+:2005,will2007inferring},
there has not been a formal definition of joint folding and alignment. Therefore there is a need to
develop such a formulation to provide mathematical rigor to this important area.
Luckily, in the sister field of computational linguistics, there is a very similar problem
``synchronous parsing'', which jointly parses and aligns a sentence pair from two languages such as English and Chinese \cite{wu:1997}; it basically extends single-sentence parsing to two sentences, just like homologous folding extends single-sequence folding to two sequences. Synchronous parsing is rigorously formulated by synchronous context-free grammars \cite{lewis+sterns:1968,aho+ullman:1969,chiang:2007}, which extend the well-known context-free grammars from one language to two languages. So we naturally borrow this concept to formulate homologous folding. See Table~\ref{tab:scfg} for the correspondence between language parsing and RNA folding. Below we start with a quick review of context-free grammars for RNA folding.


For one RNA sequence $\mathbf{x}=x_1 x_2 \ldots x_n$ with 
each $x_i \in \{\nucA, \nucU, \nucC, \nucG\}$, 
the minimum free energy change (MFE)~\cite{mathews+turner:2006} structure $\hat{\mathbf{s}}$ is the best-scoring structure among all possible structures $\mathcal{S}(\mathbf{x})$: 
\begin{equation} \label{eq:1}
\hat{\mathbf{s}} = \argmin_{\mathbf{s} \in \mathcal{S}(\mathbf{x})}  \Delta G^{\circ}(\mathbf{x}, \mathbf{s})
\end{equation}
where 
$ \Delta G^{\circ}(\mathbf{x}, \mathbf{s})$ is the free energy of the structure $\mathbf{s}$ for the sequence $\mathbf{x}$. 
The classical solution for finding the pseudoknot-free MFE structure 
is the $O(n^3)$-time dynamic programming algorithm~\cite{nussinov+jacobson:1980, zuker+stiegler:1981}, whose search space is usually formulated by 
a context free grammar (CFG). 
Formally, 
a CFG is 4-tuple $G=\langle V, \Sigma, R, S\rangle$, 
where $V$ is the set of nonterminals, 
$\Sigma$ is the set of terminals ($\Sigma$ = \{\nucA, \nucU, \nucC, \nucG\}),
$R$ is the set of production rules, 
and $S\in V$ is the start symbol.
Each rule $r\in R$ has the form $A \rightarrow \alpha$, 
where $A \in V$ is rewritten into $\alpha \in (V \cup \Sigma)^{\ast}$ where~$^\ast$ denotes zero or more repetitions.

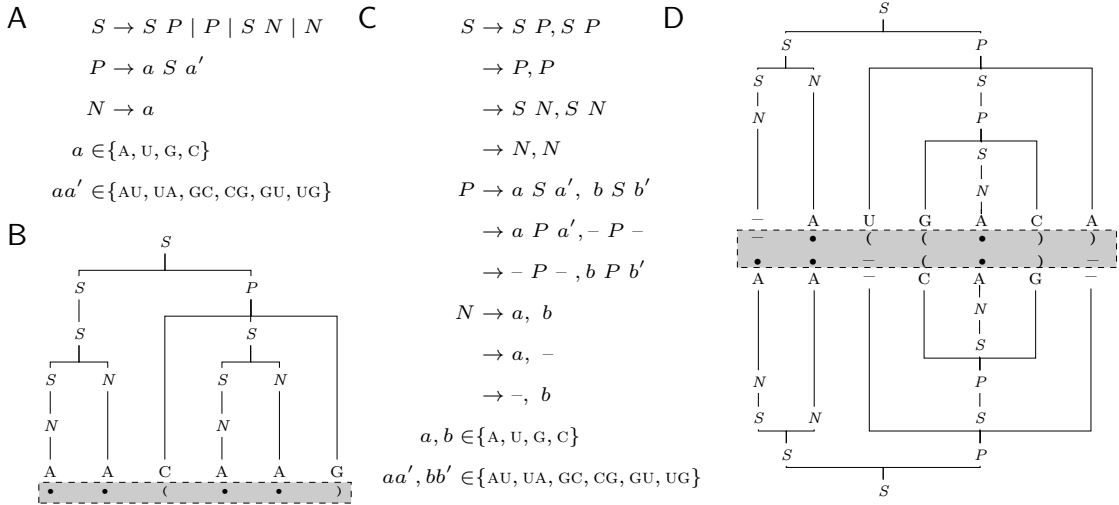
\begin{figure*}[!t]
\centering
\begin{tabular}{ccc}
\hspace{-4.5cm}{\panel{A}} &  \hspace{-4.5cm}{\panel{C}}  & \hspace{-6.5cm}{\panel{D}}\\[-.8cm] 
\raisebox{6.9cm}{
\parbox{0pt}{
    \begin{align*}
       S &\rightarrow S \ P \mid P \mid S \ N \mid N \\ 
       P &\rightarrow a \ S \ a^{\prime} \\
       N &\rightarrow a \\
       a \in & \{\nucA, \nucU, \nucG, \nucC\} \\
       aa^{\prime} \in & \{\nucA \nucU, \nucU \nucA, \nucG \nucC, \nucC\nucG,\nucG\nucU, \nucU\nucG \}    
   \end{align*}
  }}   &
\raisebox{5cm}{
\parbox{0pt}{
    \begin{align*}
       S &\rightarrow S \ P, S \ P \\
                 &\rightarrow P, P \\
                 &\rightarrow S \ N, S \ N \\
                 &\rightarrow N, N \\
       P &\rightarrow a \ S \ a^{\prime}, \ b \ S \ b^{\prime} \\
                &\rightarrow a \ P \ a^{\prime},  \gap \ P \ \gap \\
                &\rightarrow \gap \ P \ \gap \ , b \ P \ b^{\prime} \\
      N &\rightarrow a, \ b \\
                &\rightarrow a, \ \gap \\
                &\rightarrow \gap, \ b \\
         a, b \in & \{\nucA, \nucU, \nucG, \nucC\} \\
         aa^{\prime}, bb^{\prime} \in & \{\nucA \nucU, \nucU \nucA, \nucG \nucC, \nucC\nucG,\nucG\nucU, \nucU\nucG \}    
   \end{align*}
  }}   &
\resizebox{.3\textwidth}{!}{
\raisebox{2.2cm}{
\begin{tikzpicture}[sloped]
\tikzset{level distance=18pt}
\tikzset{sibling distance=15pt}
\tikzset{every tree node/.style={align=center,anchor=north}}
\begin{scope}[frontier/.style={distance from root=105pt}]
\tikzset{edge from parent/.style={draw,edge from parent path={(\tikzparentnode.south)-- +(0,-6pt)-| (\tikzchildnode)}}}
\Tree [.\it S 
               [.\it S 
                   [.\it S [.\it N {\fontsize{10.5}{0}\selectfont \gap} ] ] 
                   [.\it N {\fontsize{10.5}{0}\selectfont \nucA} ]
               ]
               [.\it P {\fontsize{10.5}{0}\selectfont \nucU}
                   [.\it S 
                       [.\it P {\fontsize{10.5}{0}\selectfont \nucG}
                           [.\it S 
                               [.\it N {\fontsize{10.5}{0}\selectfont \nucA} ]
                           ] 
                           C
                       ]
                   ] 
                   {\fontsize{10.5}{0}\selectfont \nucA}
               ] 
         ]
\end{scope}
\begin{scope}[yshift=-3.3in,grow'=up,frontier/.style={distance from root=105pt}]
\tikzset{edge from parent/.style={draw,edge from parent path={(\tikzparentnode.north)-- +(0,6pt)-| (\tikzchildnode)}}}
\Tree [.\it S 
               [.\it S 
                   [.\it S [.\it N {\fontsize{10.5}{0}\selectfont \nucA} ] ] 
                   [.\it N {\fontsize{10.5}{0}\selectfont \nucA} ]
               ]
               [.\it P {\fontsize{10.5}{0}\selectfont \gap}
                   [.\it S 
                       [.\it P {\fontsize{10.5}{0}\selectfont \nucC}
                           [.\it S 
                               [.\it N {\fontsize{10.5}{0}\selectfont \nucA} ]
                           ] 
                           {\fontsize{10.5}{0}\selectfont \nucG}
                       ]
                   ] 
                   {\fontsize{10.5}{0}\selectfont \gap}
               ] 
         ]
\end{scope}
\draw[dashed,black,fill=black!20] (-2.5,-4.7) rectangle ++(6.5,0.65);
\node(00) at (-2.15,-4.2) {\fontsize{10.5}{0}\selectfont {\gap}};
\node(11) at (-1.2,-4.2) {\fontsize{6.5}{0}\selectfont $\bullet$};
\node(22) at (-.25,-4.2) {\fontsize{6.5}{0}\selectfont {\textbf (}};
\node(33) at (0.7,-4.2) {\fontsize{6.5}{0}\selectfont {\textbf (}};
\node(44) at (1.7,-4.2) {\fontsize{6.5}{0}\selectfont $\bullet$};
\node(55) at (2.7,-4.2) {\fontsize{6.5}{0}\selectfont {\textbf )}};
\node(66) at (3.6,-4.2) {\fontsize{6.5}{0}\selectfont {\textbf )}};
\node(00) at (-2.15,-4.6) {\fontsize{6.5}{0}\selectfont $\bullet$};
\node(11) at (-1.2,-4.6) {\fontsize{6.5}{0}\selectfont $\bullet$};
\node(22) at (-.25,-4.6) {\fontsize{10.5}{0}\selectfont {\gap}};
\node(33) at (0.7,-4.6) {\fontsize{6.5}{0}\selectfont {\textbf (}};
\node(44) at (1.7,-4.6) {\fontsize{6.5}{0}\selectfont $\bullet$};
\node(55) at (2.7,-4.6) {\fontsize{6.5}{0}\selectfont {\textbf )}};
\node(66) at (3.6,-4.6) {\fontsize{10.5}{0}\selectfont {\gap}};
\end{tikzpicture}}} \\ [-5.5cm]
\hspace{-4.5cm}{\panel{B}} \\[-.3cm] 
\resizebox{.25\textwidth}{!}{
\raisebox{1.9cm}{
\begin{tikzpicture}[sloped]
\tikzset{level distance=22pt, sibling distance=15pt}
\tikzset{every tree node/.style={align=center,anchor=north}}
\begin{scope}[frontier/.style={distance from root=110pt}]
\tikzset{edge from parent/.style={draw,edge from parent path={(\tikzparentnode.south)-- +(0,-8pt)-| (\tikzchildnode)}}}
\Tree [.\it S 
               [.\it S 
                   [.\it S 
                       [.\it S [.\it N A ] ] 
                       [.\it N A ] 
                   ] 
               ]
               [.\it P C
                   [.\it S 
                       [.\it S 
                           [.\it N A ]
                       ] 
                       [.\it N A ] 
                   ] 
                   G
               ] 
         ]
\end{scope}
\draw[dashed,black,fill=black!20] (-2.1,-4.6) rectangle ++(5.2,0.35);
\node(00) at (-1.9,-4.4) {\fontsize{5}{0}\selectfont $\bullet$};
\node(11) at (-1,-4.4) {\fontsize{5}{0}\selectfont $\bullet$};
\node(22) at (0,-4.4) {\fontsize{5}{0}\selectfont {\textbf (}};
\node(33) at (1,-4.4) {\fontsize{5}{0}\selectfont $\bullet$};
\node(44) at (1.9,-4.4) {\fontsize{5}{0}\selectfont $\bullet$};
\node(55) at (2.9,-4.4) {\fontsize{5}{0}\selectfont {\textbf )}};
\end{tikzpicture}}} \\[-1cm]
\end{tabular}
\caption{({\bf A}) Context Free Grammar (CFG) formulation for parsing one sequence. 
({\bf B}) An example illustrates that CFG generates RNA sequence $\nucA\nucA\nucC\nucA\nucA\nucG$ via one possible derivation. 
The corresponding structure imposed by the derivation is annotated below the tree with gray background.  
({\bf C}) Synchronous Context Free Grammar (SCFG) formulation for simultaneously folding and aligning two sequences.
({\bf D}) And example represents that SCFG yields an aligned sequence pair (\gap\nucA\nucU\nucG\nucA\nucC\nucA, \nucA\nucA\gap\nucC\nucA\nucG\gap) via one possible derivation. 
The original sequence pair (\nucA\nucU\nucG\nucA\nucC\nucA, \nucA\nucA\nucC\nucA\nucG)  can be obtained by removing gaps directly. 
The secondary structures are shown with gray background. 
Not that we do not consider sharpturn constraint on hairpins to simplify  examples ({\bf B} and {\bf D}). 
\label{fig:scfg_formulation}
}
\end{figure*}

As an example, Fig.~\ref{fig:scfg_formulation}A shows a CFG corresponding to the
Nussinov algorithm~\cite{nussinov+jacobson:1980}.
These nonterminals represent structural components:
$S$ for an arbitrary span, 
$P$ for a span with two ends paired, and $N$ for an unpaired nucleotide.
As a shorthand notation, 
we use $a$ to represent a nucleotide, 
and $aa^{\prime}$ to represent a base pair. 
An RNA sequence $\mathbf{x}$ can be derived from the grammar $G$ by applying a series of production rules ($S \overset{\ast}{\Rightarrow}_G \mathbf{x}$).
Each derivation also implies 
a RNA secondary structure. 
Fig.~\ref{fig:scfg_formulation}B shows a derivation of $G$ 
for sequence \nucA\nucA\nucC\nucA\nucA\nucG\/ 
along with the secondary structure ``$\mydot\mydot\leftb\mydot\mydot\rightb$'' in the dot-bracket format (in gray shades,
where ``\mydot'' represents an unpaired position, and ``\leftb'' and ``\rightb'' indicate paired positions).

\smallskip


Now we extend this framework to handle two sequence folding,
by extending CFG to synchronous CFG (SCFG) \cite{chiang:2007}.
An SCFG $G^{\prime}$ is still a 4-tuple $\langle V, \Sigma^{\prime}, R^{\prime}, S\rangle$, 
where $V$ and $S$ remain unchanged, 
and the new terminal set $\Sigma^{\prime} = \{\nucA, \nucU, \nucG, \nucC, \gap\}$ 
includes a  gap symbol (\gap) for alignment. 
Each {\em synchronous production rule} in $R^{\prime}$ (see Fig.~\ref{fig:scfg_formulation}C)
now has two parts on the right hand side to capture two sequences:
$$A \rightarrow \alpha, \; \beta$$
where $A \in V$ and $\alpha, \beta \in (V \cup \Sigma)^{\ast}$. 
For example, $S \rightarrow S\ P$ is extended to $S \rightarrow S\ P, \, S\ P$.
Note that there is a one-to-one correspondence  between the nonterminals in $\alpha$ and 
the nonterminals in $\beta$.

Although the grammar $G^\prime$ requires each nonterminal (structural component) 
in one structure to correspond to another nonterminal in the other structure, 
it allows some variation on the structures of two sequences to some extent by inserting base pairs and unpaired nucleotides.
Specifically, the rule
$$P \rightarrow a\ S\ a^{\prime}, \; b\ S\ b^{\prime}$$ 
indicates that the based pairs $(a, a')$ 
and $(b, b')$ 
are aligned ($a$ with $b$ and $a'$ with $b'$).
But the rule
$$P \rightarrow a\ P\ a^{\prime},\; \gap\ P\ \gap$$ 
represents that one base pair $(a, a^\prime)$ is inserted in the first sequence
and gaps ($\gap$) are added to the second sequence for alignment. 
Similarly, 
$$P \rightarrow \gap\ P\ \gap,\; b\ P\ b^{\prime}$$ 
indicates that one base pair $(b, b^\prime)$ is inserted in the second sequence.
In addition, the productions derived from $N$ provide flexibility on the length of the corresponding unpaired regions ($N$) by inserting/deleting a nucleotide in one sequence. For example,
$$N \rightarrow a,b$$ 
aligns two unpaired (``\mydot'') nucleotides $a$ and $b$ from two sequences, while 
$$N \rightarrow a,\gap$$ 
inserts one unpaired nucleotide $a$ in the first sequence. 
Therefore, the SCFG $G^\prime$ folds two sequences with generally similar structures, 
but does not require them to be exactly the same. 
It allows freedom in the number of base pairs in corresponding helices, 
as well as the length of corresponding unpaired regions. 

More formally, a derivation of SCFG, notated $S \overset{\ast}{\Rightarrow}_{G'} \langle \bar{\vecx}, \bar{\vecy}\rangle$, generates 
a pair of {\em aligned sequences},
along with one secondary structure for each sequence.
Fig.~\ref{fig:scfg_formulation}D
demonstrates one such derivation
that generates the aligned sequence pair:
\begin{center}
\begin{BVerbatim}
-.((.))
-AUGACA
AA-CAG-
..-(.)-
\end{BVerbatim}
\end{center}
where both the sequences and structures are aligned by inserting gaps (\gap).
The original sequences and structures can be obtained by removing gaps.

\subsection{Integrating HMM-based Alignment Model}\label{subsec2}
For a sequence pair $\langle \vecx, \vecy\rangle =  \langle x_1 x_2 \ldots x_{n_1},\, y_1 y_2 \ldots y_{n_2}\rangle$ with sequence length $n_1$ and $n_2$, respectively, 
we denote a possible alignment $\mathbf{a}$ of two equal-length sequences with gaps,
$\langle x^{\prime}_{1}x^{\prime}_{2}\ldots x^{\prime}_{m}, \, y^{\prime}_{1}y^{\prime}_{2} \ldots y^{\prime}_{m}\rangle$ 
with the same sequence length $m$  $(m\geq\max(n_1, n_2))$ by inserting gaps in two sequences, 
thus $x^{\prime}_{i}, y^{\prime}_{j} \in \{\nucA, \nucU, \nucC, \nucG, \gap\}$. 
Naturally, 
with the same sequence length, 
the alignment can be treated as a sequence of pairs $\langle x^{\prime}_{1},\, y^{\prime}_{1}\rangle, \langle x^{\prime}_{2},\, y^{\prime}_{2} \rangle, \ldots , \langle x^{\prime}_{m},\, y^{\prime}_{m}\rangle$. 
We use a Hidden Markov Model to model the pairwise alignment,
which consists of three hidden states: 
$\nearrow$, $\rightarrow$ and $\uparrow$ representing alignment of two nucleotides, 
inserting one nucleotide in the first sequence, and inserting one nucleotide in the second sequence, respectively. 
Correspondingly, 
the emission/observation is $\langle x_i^{\prime}, \, y_j^{\prime}\rangle$, $\langle x_i^{\prime}, \gap\rangle$ and $\langle \gap, y_j^{\prime}\rangle$, respectively, 
where $x_i^{\prime}$ and $y_j^{\prime}$ are nucleotides rather than gaps. 
The Viterbi alignment path is the most likely sequence of hidden states to generate two sequences (ignoring gaps) among all possible alignment paths $\mathcal{A}(\mathbf{x}, \mathbf{y})$:
\begin{equation} \label{eq:2}
\begin{split}
\hat{\mathbf{a}} =& \argmin_{\mathbf{a} \in \mathcal{A}(\mathbf{x}, \mathbf{y})}  p (\mathbf{a},  \mathbf{x}, \mathbf{y}) \\ 
=&  \argmin_{\mathbf{a} \in \mathcal{A}(\mathbf{x}, \mathbf{y})}  \Pi_{i=1}^{m} {p}_{\text{t}}(h_i \mid h_{i-1}) {p}_{\text{t}}( \langle x^{\prime}_i, y^{\prime}_i \rangle \mid h_i)
\end{split}
\end{equation}
where $h_i \in \{\nearrow, \rightarrow, \uparrow \}$ is the hidden state, starting from $h_0=\nearrow$, 
and
${p}_{\text{t}}(h_i \mid h_{i-1})$ and ${p}_{\text{e}}( \langle x^{\prime}_i, y^{\prime}_i) \mid h_i \rangle$ are the transition  and the emission probabilities, respectively. 

To formalize the integrated Sankoff+HMM framework,
we need to explicitly generate structures and the alignment state sequence,
so we further extend the 2-component SCFG $G^\prime$ to 
a 5-component SCFG $G^{\prime\prime} = \langle V, \Sigma ^{\prime\prime}, R^{\prime\prime}, S \rangle$.\footnote{Such use of SCFG to explicitly model structures
is also found in natural language, e.g., between syntax and semantics \cite{shieber+schabes:1990}.}
The terminal set is extended to $\{\nucA, \nucU, \nucG, \nucC, \mydot, \leftb, \rightb, \nearrow, \rightarrow, \uparrow\}$,
where ``\mydot'', ``\leftb'' and ``\rightb'' represent structures,
and $\nearrow$, $\rightarrow$ and $\uparrow$ represent alignment states
(note that gap \gap\/ is no longer needed).  
The production rules are further extended to have five parts on the right side: 
$$A \rightarrow \alpha, \beta, \alpha^{\prime}, \beta^\prime, \theta$$
where $A \in V$, 
$\alpha, \beta \in (V \cup \{\nucA, \nucU, \nucG, \nucC\})^{\ast}$, 
$\alpha^{\prime}, \beta^{\prime} \in (V\cup\{\mydot, \leftb, \rightb\})^{\ast}$, 
and $\theta\in (V\cup\{\nearrow, \rightarrow, \uparrow\})^{\ast}$. 
For example, 
we extend the ``aligned pair'' rule $P \rightarrow a\ P \ a^\prime,\, b\ P \ b^\prime$ to 
$$P \rightarrow a\ P\ a^\prime, \, b\ P\ b^\prime,\, \leftb\ P\ \rightb,\, \leftb\ P\ \rightb,\, \nearrow P \nearrow$$
and the ``pair-gap'' rule 
$P \rightarrow a\ P\ a^\prime,\, \gap\ P\ \gap$ to 
$$P \rightarrow a\ P\ a^\prime,\,  P, \, \leftb\ P\ \rightb,\,  P ,\, \rightarrow P \rightarrow$$
where we remove the gaps in the sequences and structures.
The ``aligned unpaired'' rule
$N \rightarrow a, b$ becomes
$$N \rightarrow a, \, b,\, \mydot,\,  \mydot,\, \nearrow$$
and the ``unpaired-gap'' rule $N \rightarrow \gap, b$ becomes
$$N \rightarrow \epsilon, \, b,\, \epsilon,\,  \mydot,\, \uparrow$$
where $\epsilon$ denotes empty string.

Now a derivation in $G''$ 
generates 
a 5-tuple:
$$S \overset{\ast}{\Rightarrow}_{G^{\prime\prime}} \langle \mathbf{x}, \mathbf{y}, \mathbf{s_x}, \mathbf{s_y}, \mathbf{a} \rangle$$ 
where $\mathbf{x}$ and $\mathbf{y}$ are the two input sequences (without gaps),
$\mathbf{s_x}$ and $\mathbf{s_y}$ are their corresponding secondary structures, 
and $\mathbf{a}$ is the sequence of alignment hidden states. 

\begin{figure*}[t!]
\center
\begin{tabular}{c}
\includegraphics[width=\linewidth, trim={80 0 80 0}, clip]{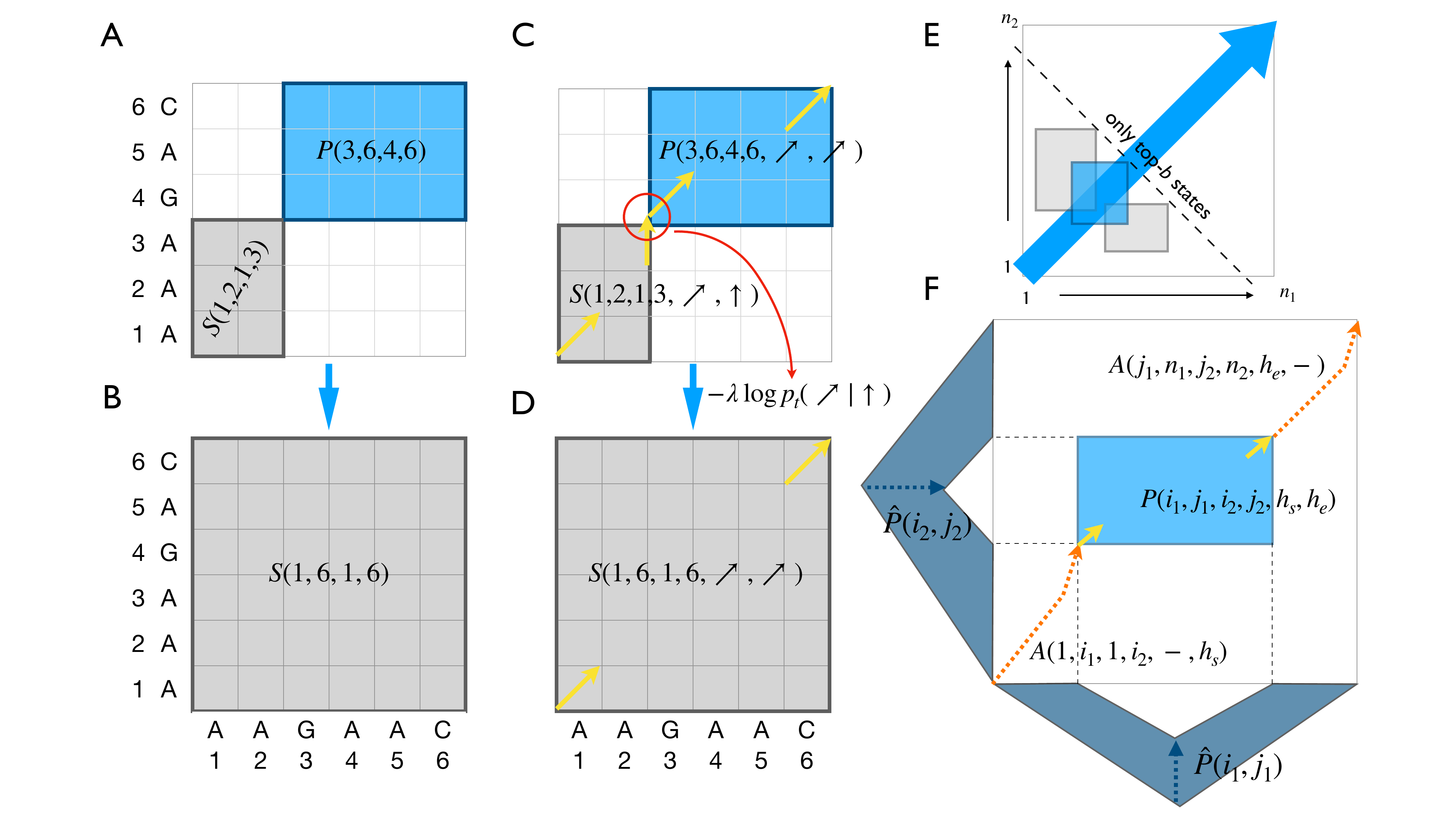}
\end{tabular}
\hfill
\caption[\lso Contributions]{
({\bf A--B}) Deductive system of Nussinov model with a simple alignment model for simultaneously folding and alignment of two sequences. Concatenate two adjacent states $S(1, 2, 1, 3)$ and $P(3, 6, 4, 6)$ ({\bf A}) to $S(1, 6, 1, 6)$ ({\bf B}). 
({\bf C--D}) Deductive system of Nussinov model with a HMM-based alignment model for simultaneously folding and alignment of two sequences. Two more dimensions are added in the states to indicates alignment hidden state of start and end positions. 
Concatenate two adjacent state $S(1, 2, 1, 3, \nearrow, \uparrow)$ and $P(3, 6, 4, 6, \nearrow, \nearrow)$ ({\bf C}) to $S(1, 6, 1, 6, \nearrow, \nearrow)$ ({\bf D}). 
The solid yellow arrows show alignment hidden states of start and end positions. 
({\bf E}) \lso's computation along the diagonal (from bottom left to top right) makes it possible to further apply the beam pruning heuristic~\cite{huang+sagae:2010}. 
({\bf F}) \lso applies the A$^{\star}$ algorithm during beam pruning to speed up searching. 
The admissible heuristic values includes $\hat{P}(i_1, j_1)$, $\hat{P}(i_2, j_2)$, $A(1, i_1, 1,  i_2,  -, h_s)$ and $A(j_1, n_1,  j_2,  n_2, h_e, -)$. 
$\hat{P}(i_1, j_1)$ is the minimum free energy change of folding regions $x_1\ldots x_{i_1-1}$ and $x_{j_1+1} \ldots x_{n_1}$ of sequence $\mathbf{x}$ conditioned on $(x_{i_1}, x_{j_1})$ forming a base pair. 
This score is obtained by folding $\mathbf{x}$ separately from pre-processing. 
A similar definition applies to $\hat{P}(i_2, j_2)$. 
$A(1, i_1, 1,  i_2,  -, h_s)$ is the Viterbi alignment path of $x_1\ldots x_{i_1}$ and $y_1\ldots y_{i_2}$ with constrained alignment state $h_s$ imposed on $(x_{i_1}, y_{i_2})$. 
This probability can be computed in pre-processing. 
A similar idea applies to $A(j_1, n_1,  j_2,  n_2, h_e, -)$. 
\label{fig:algs}}
\end{figure*}

Given two RNA homologous sequences $\mathbf{x}$ and $\mathbf{y}$, 
and a synchronous context free grammar $G''$, 
the goal of simultaneously folding and alignment of RNA sequences is to find the most likely derivation tree,
i.e., secondary structures $\mathbf{s}_x$ and $\mathbf{s}_y$,  
to generate an alignment $\mathbf{a}$ of $\mathbf{x}$ and $\mathbf{y}$ with the minimum weighted sum of folding and alignment cost:
\begin{equation}  \label{eq:3}
\min_{S \overset{\ast}{\Rightarrow}_{G^{\prime\prime}} \langle \mathbf{x}, \mathbf{y}, \mathbf{s_x}, \mathbf{s_y}, \mathbf{a} \rangle} \left [ \Delta G^{\circ} (\mathbf{x}, \mathbf{s}_x) + \Delta G^{\circ} (\mathbf{y}, \mathbf{s}_y) - \lambda \log {p} (\mathbf{a},  \mathbf{x}, \mathbf{y})\right ] 
\end{equation}

There is a trade-off between free energy changes ($\Delta G^{\circ} (\mathbf{x}, \mathbf{s}_1) + \Delta G^{\circ} (\mathbf{y}, \mathbf{s}_2)$) and the alignment cost ($\log{p} (\mathbf{a}, \mathbf{x}_1, \mathbf{x}_2)$),
which is balanced by the hyperparameter $\lambda$. 
In the complete model, 
$\Delta G^{\circ} (\mathbf{x}, \mathbf{s}_1)$ and $\Delta G^{\circ} (\mathbf{y}, \mathbf{s}_2)$ are calculated using loop-based Turner free-energy model~\cite{mathews+:1999,Mathews+:2004}, 
and ${p} (\mathbf{a}, \mathbf{x}_1, \mathbf{x}_2)$ is estimated based on the trained HMM parameters~\cite{harmanci:+2007}. 



\section{Efficient Algorithms and Implementation}

\subsection{Dynamic Programming}\label{subsec3}
Using Nussinov algorithm as an example, 
we illustrate the deductive system of \lso in Fig.~\ref{fig:algs}A--D.

With a simple alignment model, e.g., Needleman–Wunsch~\cite{needleman1970general}, 
whose alignment states are independent with neighbors, 
states $S(i_1, j_1, i_2, j_2)$ and $P(i_1, j_1, i_2, j_2)$ is the minimum cost of simultaneously folding and alignment of two spans $x_{i_1}x_{i_1+1}\ldots x_{j_1}$ and $y_{i_2}y_{i_2+1}\ldots y_{j_2}$ from two sequences, respectively. 
$P(i_1, j_1, i_2, j_2)$ requires at least one sequence forms a base pair at the two ends of span, 
either $(x_{i_1},x_{j_1})$ or $(y_{i_2}, y_{j_2})$ forms a base pair, or both. 
As shown in Fig.~\ref{fig:algs}A--B, 
concatenating two adjacent states just sums the cost of two states directly. 

The HMM-based alignment model is more complicated due to the alignment state is the current alignment state is dependent on the previous state. 
Therefore, the states $S(i_1, j_1, i_2, j_2, h_s, h_e)$ and $P(i_1, j_1, i_2, j_2, h_s, h_e)$ are extended with two more dimensions $h_s$ and $h_e$ to indicate the alignment state of start position $(x_{i_1}, y_{i_2})$ and end position $(x_{j_1}, y_{j_2})$. 
Fig.~\ref{fig:algs}C-D use solid yellow arrows to represent alignment states $h_s$ and $h_e$ of each state. 
The dotted yellow arrows are possible alignments insides each state. 
When two states are concatenated, e.g., $S(1, 2, 1, 3, \nearrow, \uparrow)$ and $P(3,6,4,6,\nearrow,\nearrow)$, 
the free energy change of secondary structure can be added.
However, the final alignment cost is the product of two probabilities of alignment path and a transition probability from $\uparrow $ to $\nearrow$. 
To get a larger $S$ by concatenating two small states $S$ and $P$, 
the state $S(1, 6, 1, 6, \nearrow, \nearrow)$ only keeps $h_s$ ($\nearrow$) from $S(1, 2, 1, 3, \nearrow, \rightarrow)$ and $h_e$ ($\nearrow$) from $P(3,6,4,6,\nearrow,\uparrow)$ and ignores the intermediate alignment states (Fig.~\ref{fig:algs}D). 

\subsection{Linearization} \label{subsec4}
Inspired by \linearfold~\cite{huang+:2019}, 
the linear-time algorithm for single RNA sequence folding,  
we generalize the beam search heuristic from single-sequence folding to simultaneously folding two sequences to achieve linear runtime against the sum of sequence lengths.
\lso parses two RNA sequences along diagonal (from bottom left to top right) (see Fig.~\ref{fig:algs}E).
Although, the current version of algorithm still runs in $O(n^6)$ time for two sequences,  
the diagonal direction allows us to further employ beam pruning heuristic~\cite{huang+sagae:2010,huang+:2019,zhang+:2020a}, 
which reduces to linear runtime.
More sepcifically, 
at each step $s$ ($s = 1...n_1 + n_2)$, 
for all candidates $P(i_1, j_1, i_2, j_2, h_s, h_e)$ ($j_1 + j_2 = s$),  
we only keep the $b$ top-scoring states and prune less promising ones because they are less likely to be part of the optimal final results. 
This results in an approximate search algorithm in $O(nb^2)$ time.


\subsection{A$^{*}$ Algorithm} \label{subsec5}
\lso applies the {A$^{*}$ algorithm to further accelerate searching during beam search. 
The heuristic values are from single sequence folding and sequence alignment. 
Formally, 
during beam search, 
for each step $s$ (from 1 to $n_1 + n_2$), 
and each state candidate $P(i_1, j_1, i_2, j_2, h_s, h_e)$ with $j_1 + j_2 = s$,
\lso builds a ``global" cost by adding an approximately estimated distance to the destination. 
For folding, 
\lso gets $\hat{P}(i_1, j_1)$,
which represents the minimum free energy change of folding regions $x_1\ldots x_{i_1-1}$ and $x_{j_1+1} \ldots x_{n_1}$ for sequence $\mathbf{x}$ from single sequence folding in pre-processing. 
Similarly,  \lso obtains $\hat{P}(i_2, j_2)$ as the minimum free energy change of folding regions $y_1\ldots y_{i_2-1}$ and $y_{j_2+1} \ldots y_{n_2}$ for sequence $\mathbf{y}$. 
For alignment, 
\lso looks up the probability of the Viterbi alignment of two prefix sequences $x_1\ldots x_{i_1}$ and $y_1 \ldots y_{i_2}$ as 
$A(1, i_1, 1, i_2, -, h_s)$, 
which constrains the alignment state of $(x_{i_1}, y_{i_2})$ to be $h_s$. 
\lso also pre-computes the probability of the Viterbi alignment of any two postfix sequences $x_{j_1}\ldots x_{n_1}$ and $y_{j_2} \ldots y_{n_2}$  as $A(j_1, n_1, j_2, n_2, h_e, -)$ from pre-processing, 
which limits the alignment state of $(x_{j_1}, y_{j_2})$ to be $h_e$.  

\lso sums up the free energy change of three segments ($P(i_1, j_1, i_2, j_2, h_s, h_e)$, $\hat{P}(i_1, j_1)$ and $\hat{P}(i_2, j_2)$) as a ``global'' folding score of  two whole sequences, 
and assembles probabilities of three alignment segments ($P(i_1, j_1, i_2, j_2, h_s, h_e)$, $A(1, i_1, 1, i_2, -, h_s)$ and $A(j_1, n_1, j_2, n_2, h_e, -)$)  as a ``global'' alignment score between two whole sequences, 
then computes a ``global" cost based on Equation \ref{eq:3}. 
Note that, 
only the segment $P(i_1, j_1, i_2, j_2, h_s, h_e)$ is from simultaneous folding and alignment, 
the folding costs ($\hat{P}(i_1, j_1)$ and $\hat{P}(i_2, j_2)$) and alignment costs ($A(1, i_1, 1, i_2, -, h_s) $ and $A(j_1, n_1, j_2, n_2, h_e, -)$) are independent of each other. 
\lso further applies beam search heuristic regarding ``global" costs.

\section{Results}\label{sec3}
\begin{figure}[t!]
\center
\begin{tabular}{c}
\includegraphics[width=0.8\linewidth, trim={0 0 720 0}, clip]{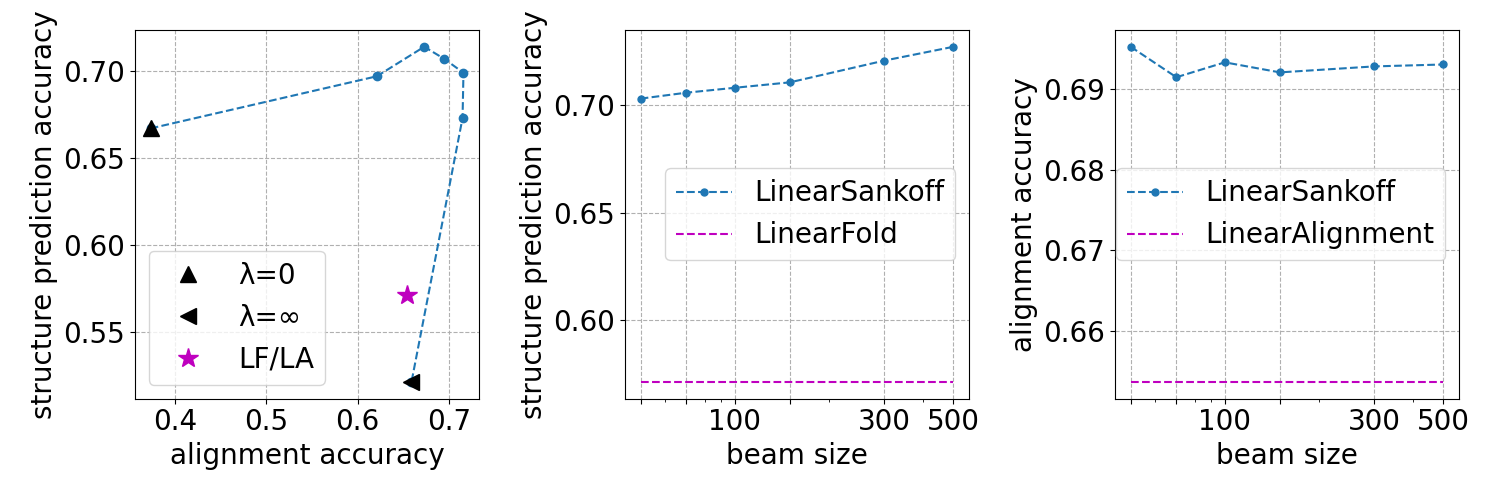}  
\end{tabular}
\hfill
\caption[Grid Search for Hyperparameters]{Grid search for the hyperparameter $\lambda$. 
Structure prediction accuracy (F1 score) against alignment accuracy (F1 score) with $\lambda$ values from 0 to $\infty$. 
LF/LA is the point that shows the structure prediction accuracy for single-sequence LinearFold calculations and alignment accuracy for LinearAlignment alignments that are not structurally informed.
\label{fig:gridsearch}}
\end{figure}

\begin{figure*}
\center
\begin{tabular}{c}
\includegraphics[width=\linewidth, trim={0 0 0 0}, clip]{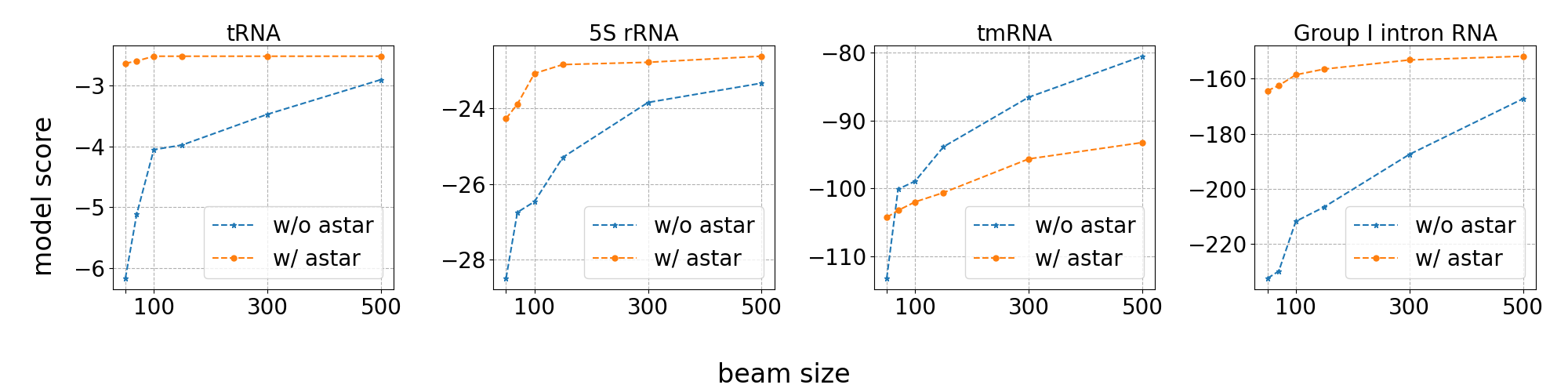}  \\ 
\end{tabular}
\hfill
\caption[Model Score with and without applying A* Search Algorithm]{Model score against beam size. The orange and blue curves represent \lso with and without applying A$^{*}$ algorithm, respectively. 
\label{fig:searchquality}}
\end{figure*}

\subsection{Hyperparameter Selection}
The weight on alignment cost ($\lambda$) is selected empirically based on performance on the training dataset. 
We benchmarked \lso with different values of $\lambda$ from 0 to infinity over four training families from \rnastralign following \turbofoldii~\cite{Tan+:2017}:
tRNA, 5S ribosomal RNA, tmRNA and Group I Intron RNA,   
and 20 sequence pairs were sampled randomly for each family. 

Fig.~\ref{fig:gridsearch} shows the secondary structure prediction accuracy (y axis) against alignment accuracy (x axis) with different values of $\lambda$. 
When $\lambda$ is 0, 
i.e., \lso barely takes advantage of alignment information (close to \dynalign), 
structure prediction accuracy of \lso is still higher than single sequence folding (\linearfold) because \lso folds two sequences to generally similar structures even though the alignment is poor ($\lambda=0$ in Fig.~\ref{fig:gridsearch}). 
Additionally, 
when $\lambda$ is infinite, 
\lso only optimizes alignment and the alignment accuracy is close to the accuracy of sequence alignment (\linearalignment, see $\lambda = \infty$ in Fig.~\ref{fig:gridsearch}).  
In between these extreme $\lambda$ values, 
as the $\lambda$ value increases, 
Fig.~\ref{fig:gridsearch} illustrates a trend that both the structure prediction and alignment accuracies first increase then decrease. 
We choose $\lambda=0.3$ as the default value which is the most closest to the top right corner. 

Fig.~\ref{fig:searchquality} compares the model scores of \lso with and without the A$^{\star}$ algorithm against the beam size over four training families. 
Both methods get higher model scores with a larger beam size. 
While \lso with A$^{\star}$ algorithm leads to have a higher model score than the plain \lso with a small beam size, e.g., 50. 
In addition to the tmRNA family, 
as the beam size increase to 500, 
\lso with the A$^{\star}$ algorithm still achieves higher model scores than the plain \lso, 
but the difference of model scores gets smaller. 
For all the results presented and discussed in the following parts are from \lso with the A$^{\star}$ algorithm. 

\subsection{Efficiency and Scalability}

To compare the runtime usage of \lso ($\lambda=0.3$ and $b=100$) and \dynalign,  
one practical implementation of  the \sankoff algorithm,  
we collected a dataset that consists of  sequence pairs from \rnastralign   
with the average sequence length ranging from 70 to 3000~\nts.
We used a Linux machine (CentOS 7.7.1908) with a 2.30 GHz Intel Xeon E5-2695 v3 CPU and 755 GB memory,
and gcc 4.8.5 for benchmarks. 

As we discussed above, 
\dynalign takes $O(n^3m^3)$ time,
 where $m$ is the average width of alignment searching space,
which correlates with the sequence identity. 
\dynalign has two modes to decide the value of $m$.
One is to require users to specify the value of $m$, 
which is fixed along the sequence.
Another mode is to generate a valid alignment searching space adaptively based on sequence identity. 
\dynalign first computes posterior alignment probabilities using the forward-backward algorithm, 
then prunes unlikely positions by a threshold, 
which is determined by sequence identity.
\lso also has these two modes. 

\begin{figure*}[t!]
\center
\begin{tabular}{c}
\includegraphics[width=\linewidth, trim={0 250 0 300}, clip]{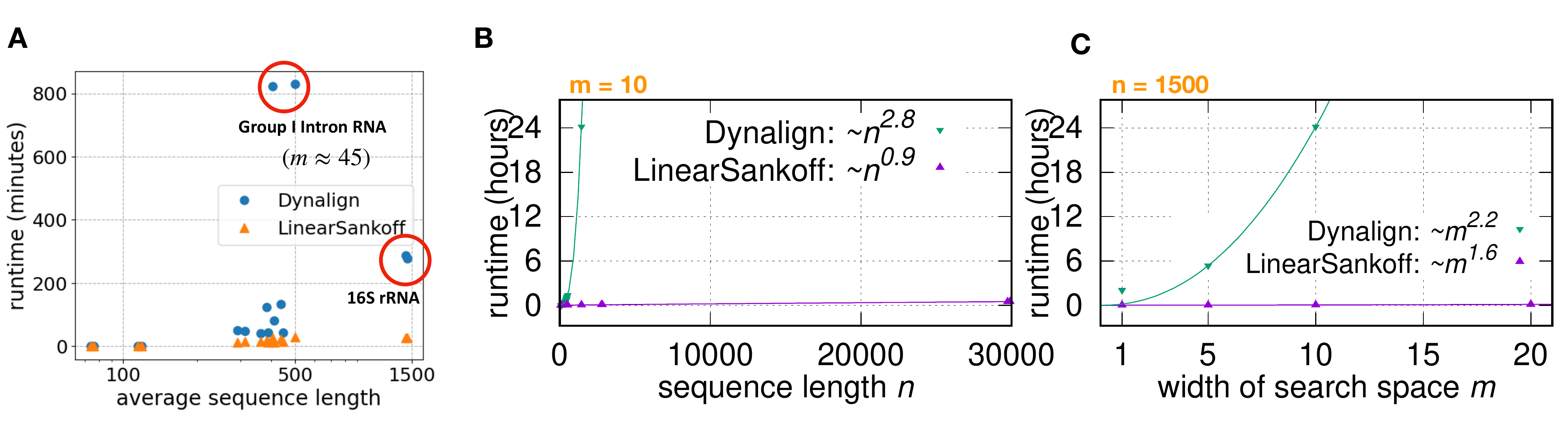}  
\end{tabular}
\hfill
\caption[Runtime Comparations]{Runtime comparisons between \dynalign and \lso ($b$=100). 
({\bf A}) The runtime of sampled pairs of sequences. Both \dynalign and \lso construct the alignment searching space adaptively based on sequence identity. 
({\bf B}) The runtime against sequence length ($n$) with the fixed width of alignment searching space ($m$=10~\nts).
({\bf C}) The runtime against the width of alignment search space ($m$) with fixed sequence length ($n$=1500~\nts).
\label{fig:runtime}}
\end{figure*}

Fig.~\ref{fig:runtime}A uses the second mode, 
which restricts the alignment searching space adaptively based on sequence identity,  
to show runtime comparison between \dynalign and \lso. 
\dynalign took more than 800 minutes for two pairs of Group I Intron sequences with sequence length $\sim$500~\nts. 
The sequence identity of these two pairs is around 0.35 with $m$ 45~\nts. 
Even though 16S rRNA sequences ($\sim$1500~\nts) are three times longer than Group I Intron sequences,
\dynalign only took one third of the runtime spent on Group I Intron sequence pairs due to the high sequence identity (0.85) and a narrow searching space ($m$ = 4~\nts) of the 16S rRNA sequence pairs. 

Thanks to the beam pruning,
although \lso performs a more complicated alignment model than \dynalign, 
\lso is significantly faster than \dynalign, 
especially for diverse or long sequences.  
In Fig.~\ref{fig:runtime}A, 
for instance, 
\dynalign took $\sim$13 hours for Group I Intron sequence pairs,
which are the most diverse sequences among sampled data,
and  $\sim$3 hours for 16S rRNA sequence pairs, 
which are the longest sequences.
While, \lso only needs 5 minutes for Group I Intron sequence pairs and 30 seconds for 16S rRNA sequence pairs, respectively. 

Both \dynalign and \lso's runtime correlate with both alignment searching space ($m$) and sequence length ($n$).
Therefore, we fixed alignment searching space (Fig.~\ref{fig:runtime}B) and sequence length (Fig.~\ref{fig:runtime}C) to show correlation with two variables. 
With a fixed $m$, 
 \dynalign scales cubically with sequence length, 
 while \lso takes linear runtime with sequence length. 
With fixed sequence length and $m$ varying from 1 to 20~\nts, 
both \dynalign and \lso scales almost quadratically with $m$.



\begin{table}[t!]
\center
\scalebox{0.8}{
\begin{tabular}{c|c|c|c|c|c}
 &  \begin{tabular}[c]{@{}c@{}}SRP \\ RNA\end{tabular}  &  \begin{tabular}[c]{@{}c@{}}RNaseP \\ RNA\end{tabular} & \begin{tabular}[c]{@{}c@{}}telomerase \\ RNA\end{tabular}  & \begin{tabular}[c]{@{}c@{}}16S \\ rRNA\end{tabular} & overall \\ 
 \hline
sequence  length & 286 & 370 & 455   & 1140  &  \\ 
sequence identity & 0.29 & 0.48 & 0.83  & 0.85  &  \\
$m$ & 25.4 & 18.8 & 3.4  & 3.7  &  \\
\hline 
\multicolumn{6}{c}{Structure Prediction Accuracy (F1 score)} \\
\hline
\linearfold~\cite{huang+:2019} & 72.1 & 59.0 & 54.3  & 46.1  & 58.0 \\
\dynalign~\cite{mathews2002dynalign} & 72.5 & 69.2 & 66.4  & 56.3  & 66.2 \\
\foldalign~\cite{havgaard2007fast}  & 61.4 & 56.8 & 40.6  & 53.1  & 53.2 \\
\locarna~\cite{will2007inferring} & 70.9 & 60.0 & 61.7  & 58.6  & 63.0 \\ 
\scarna~\cite{tabei2006scarna} & 72.7 & 55.6 & 44.1  & 62.0  & 58.7 \\ 
\lturbofold~\cite{li2021linearturbofold} & 69.5 & 70.4 & 58.3  & 54.2  & 63.3 \\
MAFFT+\rnaalifold~\cite{bernhart+:2008} & 31.5 & 42.0 & 50.2  & 57.0  & 45.3 \\
\hline
\lso ($b$=100) & 75.4 & 70.4 & \bf{68.6}  & \bf{60.1}  &68.6 \\
\lso ($b$=$\infty$) & \bf{76.3} & \bf{73.3} & 67.5 & 58.8  &  \bf{69.0}\ \\
\hline
\multicolumn{6}{c}{Alignment Accuracy (F1 score)} \\
\hline
\mafft~\cite{katoh+standley:2013} & 44.4 & 70.1 &  93.1 & 97.3  &  76.2 \\
\dynalign & 43.2 & 56.6 & 70.5  & 90.1  & 65.1 \\
\foldalign  & 51.2 & 71.0 & 92.7  & 97.2  & 78.0 \\
\locarna & \bf{54.8} & 70.7 & 92.3  & 97.3  & \bf{78.8} \\ 
\scarna & 50.2 & 70.5 & 93.0  & \bf{97.4}  & 77.8 \\ 
\lturbofold & 50.8 & 69.0 & \bf{93.4}  & 97.3  & 77.6 \\
\hline
\lso ($b$=100) & 50.8 & \bf{73.0} & 91.2  & 96.8  & 77.9 \\
\lso ($b$=$\infty$) & 51.2 & \bf{73.0} &  91.3 & 96.7  &  78.1 \\
\hline
\end{tabular}
}
\hfill
\caption[Structure Prediction and Alignment Accuracies on Test Set]{Structure prediction and alignment accuracies on test set. 
\label{tab:test_acc}}
\end{table}

\subsection{Folding and Alignment Accuracies}
To evaluate \lso and several benchmarks, 
we first randomly sampled 80 sequence pairs from the other four families of the \rnastralign dataset: 
SRP RNA, telomerase RNA, RNase P RNA and 16S rRNA.
The first three rows in Tab.~\ref{tab:test_acc} summarize the basic information of these four families including the average sequence length, sequence identity, and the average alignment searching space ($m$). 
The benchmarks consist of \linearfold, \mafft, \dynalign, \foldalign, \locarna, \scarna, \lturbofold and \rnaalifold, 
which are selected from several perspectives. 
\linearfold predicts structures for a single sequence, 
and \mafft performs alignment only based on nucleotides. 
\dynalign, \foldalign, \locarna and \scarna are representative implementations of the \sankoff algorithm. 
As a workaround of the \sankoff algorithm, 
\lturbofold iteratively performs folding and alignment modules to avoid strictly simultaneous computation. 
\mafft + \rnaalifold divides the task of simultaneous folding and alignment into two consecutive independent tasks: first aligning sequences then folding the alignment. 
We ran all the tools with default settings, 
only ran \scarna with ``-rfold" mode. 

For secondary structure prediction, 
\lso (both $b$=100 and $\infty$) first perform better than single-sequence folding (\linearfold). 
\lso (both $b$=100 and $\infty$) achieves higher accuracy than Sankoff-style methods  including \dynalign, 
\foldalign, \locarna and \scarna on every test family. 
With infinite beam size, 
\lso leads to better performance on the SRP RNA and RNase P RNA families than beam size 100.  
These two families have relatively low sequence identity with large alignment searching space ($m$) among four test families, thus need a large beam size. 

Regarding alignment accuracy, 
\dynalign obtains the lowest accuracy among all benchmarks on all four test families, 
which does not include terms for sequence identity. 
\foldalign, \scarna, \lturbofold achieve comparable alignment accuracy to \lso (both $b$=100 and $\infty$). 
\lso with infinite beam size is the second-best tool in terms of alignment quality, 
and its accuracy is only lower than \locarna. 
While, 
\locarna performs worse on secondary structure prediction than \dynalign, \lturbofold and \lso (both $b$=100 and $\infty$).


\begin{figure}[t!]
\hspace{-3.9cm}{\panel{A}} \\[-0.3cm]
\centering\resizebox{\columnwidth}{!}{%
\begin{tabular}{c|ccc|ccc}
& \multicolumn{3}{c}{tRNA (4- vs. 5-branches)} & \multicolumn{3}{c}{SRP RNA (2- vs. 3-branches)} \\
& PPV & sensitivity & F1 & PPV & sensitivity & F1 \\
\hline 
\dynalign & 71.4 & 70.8 & 71.1 & 59.3 & 59.2 & 59.2 \\
\lso & 87.6 & 82.6 & 85.0 & 43.4 & 37.1 & 40.0 \\
\hline
\dynalignii & 82.7 & 84.3 & 83.5 & 70.4 & 72.9 & 71.6 \\
\lso$^{\dagger}$ & 91.6 & 91.6 & 91.6 & 73.7 & 71.3 & 72.5  \\
\hline
\end{tabular}
}
\begin{tabular}{c}
\\[-0.3cm]
\hspace{-8.2cm}{\panel{B}} \\
\includegraphics[width=\linewidth, trim={60 0 0 0}, clip]{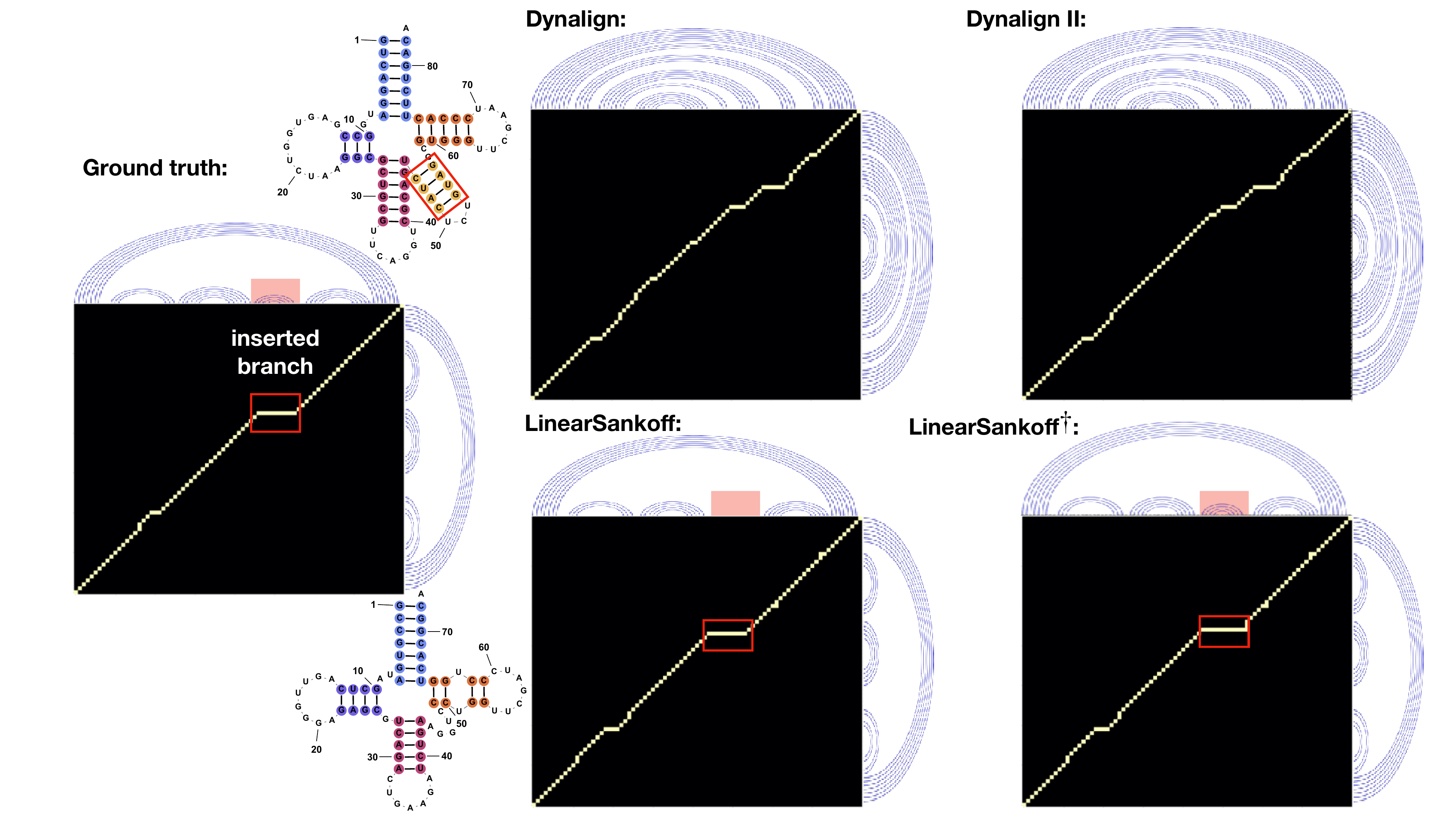}
\end{tabular}
\caption[Domain Insertion Accuracies]{({\bf A}) Structure prediction accuracies on families with domain insertion. 
Both \dynalign and \lso are unable to capture the inserted branch, 
while \dynalignii and \lso$^{\dagger}$ have the advanced domain insertion feature. 
({\bf B}) The ground truth and the predictions of \dynalign, \dynalignii, \lso and \lso$^{\dagger}$ for a sequence pair (tdbD00005111 and tdbD00001770) from the tRNA family. 
\label{fig:domain_insertion}}
\end{figure}

\subsection{Domain Insertion}
Most of implementations and variations of the \sankoff algorithm (\dynalign, \foldalign, \locarna, \scarna and \lso) fold RNA homologous sequences to generally similar structures, 
i.e., a branch in one structure must have a corresponding branch in the other structure. 
However, 
it is not guaranteed that the one-to-one correspondence always exists.  
For instance, 
some structures contains insertion or deletion of a whole branch. 
As the ground truth shown in Fig.~\ref{fig:domain_insertion}B, 
one structure (on the top of the matrix) contains one more branch (covered by a red box) than the other structure on the right side. 
The yellow curve in the black matrix represents the alignment between two sequences with a continuous long insertion corresponds to insertion of a whole branch into the top structure. 
\dynalignii~\cite{fu2014dynalign} extends the \dynalign to model domain insertion.
Following \dynalignii, 
\lso$^{\dagger}$  is able to model inserted branches as well.

To evaluate the modeling ability of \lso$^{\dagger}$,
we collected a specific dataset by sampling sequence pairs from tRNA and SRP RNA families. 
From the tRNA family, we sampled sequence pairs, whose structures consist of four branches and five branches, respectively. 
For the SRP RNA family, we sampled sequence pairs from two subfamilies (archael and long bacterial). 
Compared to the structure of the archael,
one branch is deleted from a three-branch multiloop in the structure of the long bacterial. 
Fig.~\ref{fig:domain_insertion}A shows performance of \dynalign, \dynalignii, \lso and \lso$^{\dagger}$ on two families. 
Overall, with the help of HMM-based alignment model, 
\lso$^{\dagger}$  achieves higher structure prediction accuracy than \dynalignii after adding the feature of domain insertion.

Without the feature to model domain insertion,
\lso achieves higher accuracy than \dynalign on the  tRNA  family due to the powerful HMM-based alignment model.
Although it is out of the scope of \lso to predict insertion of a whole branch, 
the HMM-based alignment model captures the signal from sequences and \lso just leaves the corresponding region unpaired (see red boxes in  \lso prediction in Fig.~\ref{fig:domain_insertion}B). 
\lso even obtains higher accuracy than \dynalignii. 
We observed that \dynalignii predicts same structures as \dynalign for some tRNA sequence pairs (see \dynalign and \dynalignii predictions in Fig.~\ref{fig:domain_insertion}B),
which is highly because the default penalty for domain insertion is relatively large for tRNA sequences (only $\sim$70~\nts). 
In other words, the free energy change of adding a new branch can not make up for the penalty of domain insertion. 
While \lso does not reply on any penalty for domain insertion, only the probability of the alignment path. 

\begin{figure*}[t]
\center
\begin{tabular}{c}
\includegraphics[width=\linewidth, trim={0 250 0 250}, clip]{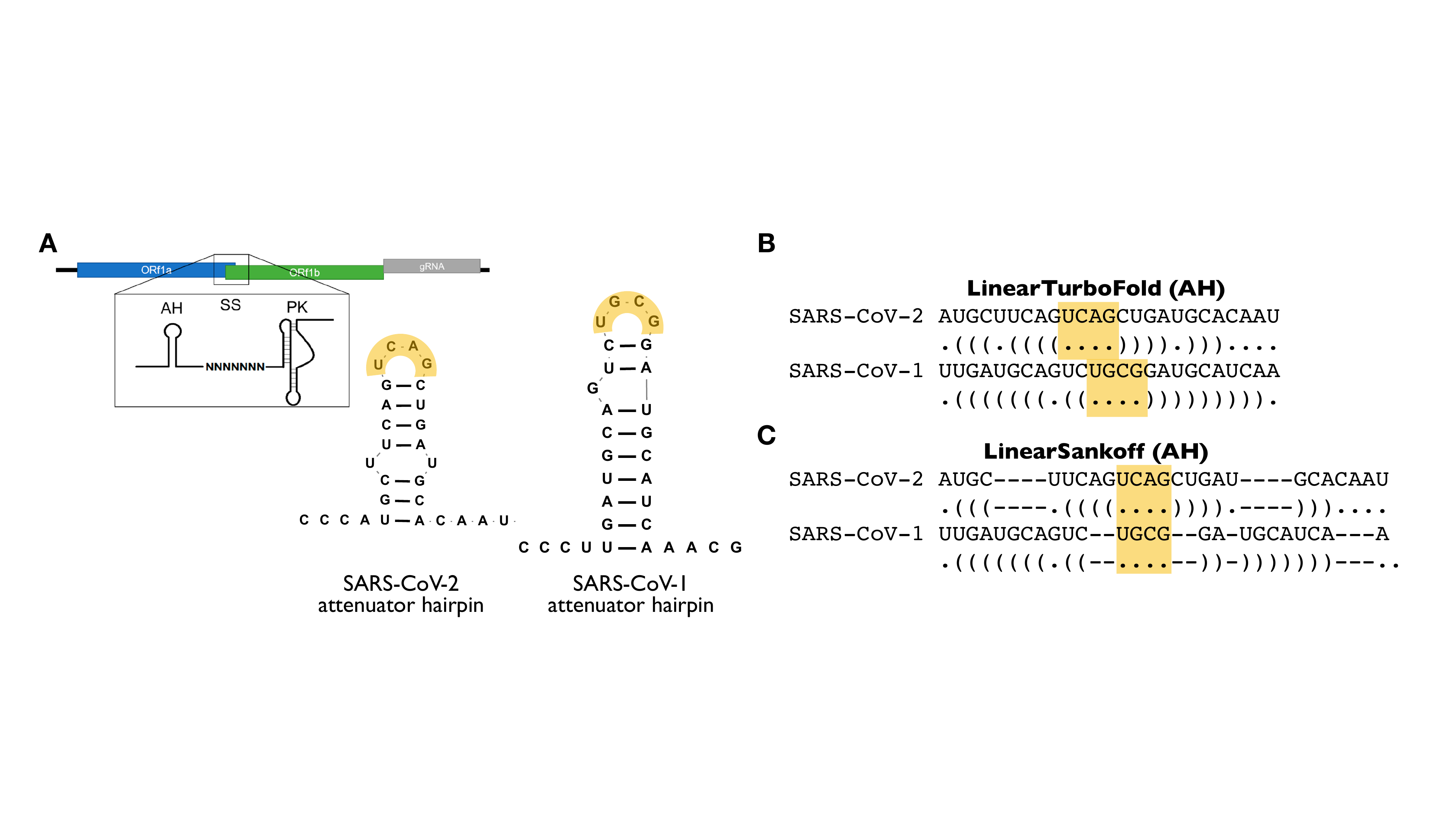}  
\end{tabular}
\hfill
\caption[Alignment of Attenuator Hairpins of SARS-CoV-2 and SARS-related Genomes]{The alignment of attenuator hairpins in the frameshifting element (FSE). 
({\bf A}) Canonical structures of attenuator hairpins in SARS-CoV-2 and SARS-CoV-1 frameshifting element. 
({\bf B}) Alignment of attenuator hairpins generated by \lturbofold over 25 SARS-CoV-2 and SARS-related genomes. 
({\bf C}) Alignment of attenuator hairpins generated by \lso. 
Clearly, the alignment from \lso is more accurate than the output of \lturbofold for the attenuator hairpins. 
\label{fig:cov}}
\end{figure*}

\subsection{Application to SARS-CoV-2 genomes}

We further applied \lso to the SARS-CoV-2 reference genome (NC\_045512.2) with the SARS-CoV-1 reference genome (NC\_004718). 
The attenuator hairpin (AH) in the frameshifting element (FSE) are conserved among SARS-CoV-2  and SARS-related genomes  
and its structures are well established as shown in Fig~\ref{fig:cov}A.  
However, \lturbofold cannot align these two attenuator hairpins from SARS-CoV-2 and SARS-CoV-1 correctly due to some extent of disagreement between folding and alignment (Fig~\ref{fig:cov}B). 
Thanks to the strong coupling between folding and alignment in \lso, 
Fig~\ref{fig:cov}C shows \lso aligns two structures properly, 
and we can  further extract conserved structures directly based on \lso's prediction without any extra manual work.

%

\section{Conclusion}

We focus on simultaneous folding and alignment of RNA homologous sequences. 
Formally, 
we borrowed synchronous context free grammars from computational linguistics to formulate homologous folding of two RNA sequences. 
We proposed \lso, 
which enhances the modeling capacity of the \sankoff algorithm by integrating it with an HMM-based alignment model. 
We devised a dynamic programming algorithm tailored to this combined Sankoff+HMM model.
In addition, 
\lso generalizes beam search heuristic from single-sequence folding to parsing two sequences simultaneously, 
which make its runtime scale linearly with  sequence length. 
\lso further applies A$^\star$ algorithm to conduct more efficient searching together with beam pruning. 
Based on evaluation on four test families and comparison with a variety of benchmarks, 
\lso achieves significantly better secondary structure accuracy than other benchmarks, 
and comparable alignment accuracy to most of the Sankoff-style tools. 
\lso is also the first joint folding and alignment algorithm to scale to full-length SARS-CoV-2 genomes,
and outperforms other tools in identifying crucial conserved structures between SARS-CoV-2 and SARS-CoV-1.

\lso is in principle extendable to multiple sequences. with several solutions. 
One option is to replace \dynalign with \lso in \multilign~\cite{Xu+:2011}, 
which progressively constructs a conserved structure to multiple sequences by conducting pairwise alignment using \dynalign. 
Another option is to
generalize \lso to take not only single sequences but also multiple sequence alignments (MSA) as input, 
i.e., simultaneously folding and alignment of MSAs. 
With such generalizations, 
\lso can progressively build the MSA along a phylogenetic tree. 
We leave these endeavors to future work.


%
%
%

\bibliographystyle{unsrt}
\bibliography{main}

\begin{thebibliography}{10}

\bibitem{Eddy:2001}
S.~R. Eddy.
\newblock Non-coding {RNA} genes and the modern {RNA} world.
\newblock {\em Nature Reviews Genetics}, 2(12):919--929, 2001.

\bibitem{Doudna+Cech:2002}
Jennifer~A. Doudna and Thomas~R. Cech.
\newblock The chemical repertoire of natural ribozymes.
\newblock {\em Nature}, 418(6894):222--228, 2002.

\bibitem{Sankoff:1985}
David Sankoff.
\newblock Simultaneous solution of the {RNA} folding, alignment and
  protosequence problems.
\newblock {\em {SIAM} Journal on Applied Mathematics}, 45(5):810–--825, 1985.

\bibitem{mathews2002dynalign}
David~H Mathews and Douglas~H Turner.
\newblock Dynalign: an algorithm for finding the secondary structure common to
  two {RNA} sequences.
\newblock {\em Journal of molecular biology}, 317(2):191--203, 2002.

\bibitem{harmanci:+2007}
Arif~Ozgun Harmanci, Gaurav Sharma, and David~H Mathews.
\newblock {Efficient pairwise {RNA} structure prediction using probabilistic
  alignment constraints in Dynalign}.
\newblock {\em BMC Bioinformatics}, 8(1):130, 2007.

\bibitem{fu2014dynalign}
Yinghan Fu, Gaurav Sharma, and David~H Mathews.
\newblock Dynalign ii: common secondary structure prediction for rna homologs
  with domain insertions.
\newblock {\em Nucleic acids research}, 42(22):13939--13948, 2014.

\bibitem{havgaard2007fast}
Jakob~H Havgaard, Elfar Torarinsson, and Jan Gorodkin.
\newblock Fast pairwise structural rna alignments by pruning of the dynamical
  programming matrix.
\newblock {\em PLOS computational biology}, 3(10):e193, 2007.

\bibitem{do+:2005}
Chuong~B Do, Mahathi~SP Mahabhashyam, Michael Brudno, and Serafim Batzoglou.
\newblock Probcons: Probabilistic consistency-based multiple sequence
  alignment.
\newblock {\em Genome research}, 15(2):330--340, 2005.

\bibitem{will2007inferring}
Sebastian Will, Kristin Reiche, Ivo~L Hofacker, Peter~F Stadler, and Rolf
  Backofen.
\newblock Inferring noncoding rna families and classes by means of genome-scale
  structure-based clustering.
\newblock {\em PLoS computational biology}, 3(4):e65, 2007.

\bibitem{tabei2006scarna}
Yasuo Tabei, Koji Tsuda, Taishin Kin, and Kiyoshi Asai.
\newblock {SCARNA}: fast and accurate structural alignment of {RNA} sequences
  by matching fixed-length stem fragments.
\newblock {\em Bioinformatics}, 22(14):1723--1729, 2006.

\bibitem{harmanci2008parts}
Arif~Ozgun Harmanci, Gaurav Sharma, and David~H Mathews.
\newblock Parts: probabilistic alignment for rna joint secondary structure
  prediction.
\newblock {\em Nucleic acids research}, 36(7):2406--2417, 2008.

\bibitem{Tabei+:2008}
Yasuo Tabei, Hisanori Kiryu, Taishin Kin, and Kiyoshi Asai.
\newblock A fast structural multiple alignment method for long {RNA} sequences.
\newblock {\em {BMC} Bioinformatics}, 9(1):33, 2008.

\bibitem{needleman1970general}
Saul~B Needleman and Christian~D Wunsch.
\newblock A general method applicable to the search for similarities in the
  amino acid sequence of two proteins.
\newblock {\em Journal of molecular biology}, 48(3):443--453, 1970.

\bibitem{durbin1998biological}
Richard Durbin, Sean~R Eddy, Anders Krogh, and Graeme Mitchison.
\newblock {\em Biological sequence analysis: probabilistic models of proteins
  and nucleic acids}.
\newblock Cambridge university press, 1998.

\bibitem{Harmanci+:2011}
Arif~O Harmanci, Gaurav Sharma, and David~H Mathews.
\newblock Turbofold: iterative probabilistic estimation of secondary structures
  for multiple {RNA} sequences.
\newblock {\em BMC bioinformatics}, 12(1):108, 2011.

\bibitem{huang+:2019}
Liang Huang, He~Zhang, Dezhong Deng, Kai Zhao, Kaibo Liu, David Hendrix, and
  David Mathews.
\newblock {LinearFold: linear-time approximate {RNA} folding by 5'-to-3'
  dynamic programming and beam search}.
\newblock {\em Bioinformatics}, 35(14):i295--i304, 07 2019.

\bibitem{hart1968formal}
Peter~E Hart, Nils~J Nilsson, and Bertram Raphael.
\newblock A formal basis for the heuristic determination of minimum cost paths.
\newblock {\em IEEE transactions on Systems Science and Cybernetics},
  4(2):100--107, 1968.

\bibitem{wu:1997}
Dekai Wu.
\newblock Stochastic inversion transduction grammars and bilingual parsing of
  parallel corpora.
\newblock {\em Computational linguistics}, 23(3):377--403, 1997.

\bibitem{lewis+sterns:1968}
Philip~M Lewis and Richard~Edwin Stearns.
\newblock Syntax-directed transduction.
\newblock {\em Journal of the ACM (JACM)}, 15(3):465--488, 1968.

\bibitem{aho+ullman:1969}
Alfred~V. Aho and Jeffrey~D. Ullman.
\newblock Syntax directed translations and the pushdown assembler.
\newblock {\em Journal of Computer and System Sciences}, 3(1):37--56, 1969.

\bibitem{chiang:2007}
David Chiang.
\newblock Hierarchical phrase-based translation.
\newblock {\em computational linguistics}, 33(2):201--228, 2007.

\bibitem{mathews+turner:2006}
David~H Mathews and Douglas~H Turner.
\newblock Prediction of {RNA} secondary structure by free energy minimization.
\newblock {\em Curr. Opin. Struct. Biol.}, 16(3):270--278, 2006.

\bibitem{nussinov+jacobson:1980}
Ruth Nussinov and Ann~B Jacobson.
\newblock Fast algorithm for predicting the secondary structure of
  single-stranded {RNA}.
\newblock {\em {PNAS}}, 77(11):6309--6313, 1980.

\bibitem{zuker+stiegler:1981}
Michael Zuker and Patrick Stiegler.
\newblock Optimal computer folding of large {RNA} sequences using
  thermodynamics and auxiliary information.
\newblock {\em {NAR}}, 9(1):133--148, 1981.

\bibitem{shieber+schabes:1990}
Stuart~M. Shieber and Yves Schabes.
\newblock Synchronous {T}ree-{A}djoining {G}rammars.
\newblock In {\em {COLING} 1990 Volume 3: Papers presented to the 13th
  International Conference on Computational Linguistics}, 1990.

\bibitem{huang+sagae:2010}
Liang Huang and Kenji Sagae.
\newblock Dynamic programming for linear-time incremental parsing.
\newblock In {\em Proceedings of ACL 2010}, page 1077–1086, Uppsala, Sweden,
  2010. {ACL}.

\bibitem{mathews+:1999}
David~H Mathews, Jeffrey Sabina, Michael Zuker, and Douglas~H Turner.
\newblock Expanded sequence dependence of thermodynamic parameters improves
  prediction of {RNA} secondary structure.
\newblock {\em J. Mol. Biol.}, 288(5):911--940, 1999.

\bibitem{Mathews+:2004}
David Mathews et~al.
\newblock Incorporating chemical modification constraints into a dynamic
  programming algorithm for prediction of {RNA} secondary structure.
\newblock {\em {PNAS}}, 101(19):7287--7292, 2004.

\bibitem{zhang+:2020a}
He~Zhang, Liang Zhang, David~H Mathews, and Liang Huang.
\newblock {LinearPartition}: linear-time approximation of {RNA} folding
  partition function and base-pairing probabilities.
\newblock {\em Bioinformatics}, 36(Supplement\_1):i258--i267, 2020.

\bibitem{Tan+:2017}
Zhen Tan, Yinghan Fu, Gaurav Sharma, and David~H. Mathews.
\newblock {TurboFold II: {RNA} structural alignment and secondary structure
  prediction informed by multiple homologs}.
\newblock {\em Nucleic Acids Research}, 45(20):11570--11581, 09 2017.

\bibitem{li2021linearturbofold}
Sizhen Li, He~Zhang, Liang Zhang, Kaibo Liu, Boxiang Liu, David~H Mathews, and
  Liang Huang.
\newblock Linearturbofold: Linear-time global prediction of conserved
  structures for rna homologs with applications to sars-cov-2.
\newblock {\em Proceedings of the National Academy of Sciences},
  118(52):e2116269118, 2021.

\bibitem{bernhart+:2008}
Stephan~H Bernhart, Ivo~L Hofacker, Sebastian Will, Andreas~R Gruber, and
  Peter~F Stadler.
\newblock {RNAalifold}: improved consensus structure prediction for {RNA}
  alignments.
\newblock {\em BMC Bioinformatics}, 9(1):1--13, 2008.

\bibitem{katoh+standley:2013}
Kazutaka Katoh and Daron~M Standley.
\newblock {MAFFT} multiple sequence alignment software version 7: improvements
  in performance and usability.
\newblock {\em Molecular Biology and Evolution}, 30(4):772--780, 2013.

\bibitem{Xu+:2011}
Zhenjiang Xu and David~H Mathews.
\newblock Multilign: an algorithm to predict secondary structures conserved in
  multiple {RNA} sequences.
\newblock {\em Bioinformatics}, 27(5):626--632, 2011.

\end{thebibliography}




\end{document}